\documentclass[twocolumn,showpacs,preprintnumbers,amsmath,amssymb]{revtex4}   
\usepackage{epsfig}
\usepackage{dcolumn}   
\usepackage{bm}        
\usepackage{amssymb}
\begin{document}

\title{The impact of the medium and the jet-medium coupling on jet measurements at RHIC and LHC}
\author{Barbara Betz$^{1}$, Florian Senzel$^{1}$, Carsten Greiner$^{1}$, and Miklos Gyulassy$^{2,3}$}
\affiliation{$^1$Institut f\"ur Theoretische Physik, Goethe Universit\"at, Frankfurt am Main, Germany\\
$^2$Department of Physics, Columbia University, 538 West 120$^{\,th}$ Street, New York,
NY 10027, USA\\
$^3$Institute of Particle Physics, Central China Normal University, Wuhan, China}

\begin{abstract}
We compare a perturbative QCD-based jet-energy loss model to the measured data of the 
pion nuclear modification factor and the high-$p_T$ elliptic flow at RHIC and LHC 
energies. This jet-energy loss model (BBMG) is currently coupled to 
state-of-the-art hydrodynamic descriptions. We report on a model 
extension to medium backgrounds generated by the parton cascade BAMPS. In addition, 
we study the impact of realistic medium transverse flow fields and a jet-medium coupling 
which includes the effects of the jet energy, the temperature of the bulk medium, 
and non-equilibrium effects close to the phase transition. By contrasting the two
different background models, we point out that the description of the high-$p_T$ elliptic flow 
for a non-fluctuating medium requires to include such a jet-medium coupling 
and the transverse flow fields. While the results for both medium 
backgrounds show a remarkable similarity, there is an impact of the 
background medium and the background flow on the high-$p_T$ elliptic flow.
\end{abstract}

\date{\today}
\maketitle

\section{Introduction}

One of the open challenges in heavy-ion physics is to gain a precise understanding
of the jet-medium dynamics, the jet-medium interactions, and the jet-energy loss 
formalism. In this letter, we study the impact of the medium and the 
details of the jet-medium coupling on the pion nuclear modification factor
($R_{AA}$) and the high-$p_T$ elliptic flow ($v_2$) measured at the
Relativistic Heavy Ion Collider (RHIC) and the Large Hadron Collider (LHC)
\cite{data1,data2,data3,data4}.

We find that both the background medium and the details of the jet-medium coupling
play an important role for the simultaneous description of the $R_{AA}$ and the 
high-$p_T$ $v_2$. This simultaneous description reveals the so-called high-$p_T$ 
$v_2$-problem \cite{Betz:2014cza,Betz:2012qq}:
For various theoretical models \cite{data1,Xu,Molnar}, the high-$p_T$ elliptic 
flow below $p_T \sim 20$~GeV is about a factor of two below the measured data 
\cite{data1,data2,data3,data4}. This effect has been discussed in literature 
\cite{Betz:2014cza,Betz:2012qq,Xu,Molnar}. Recently, it has been shown by CUJET3.0 \cite{Xu:2014tda} that 
a jet-medium coupling $\kappa=\kappa(E^2,T)$ can solve the high-$p_T$ $v_2$-problem
for non-fluctuating initial conditions. This jet-medium coupling 
depends on the energy of the jet $E$, the temperature of the medium $T$, and 
non-equilibrium effects around the phase transition of $T_c \sim 160$~MeV.

In this letter, we contrast results obtained for a background medium determined 
via the viscous hydrodynamic approach VISH2+1 \cite{VISH2+1} with the parton cascade 
BAMPS \cite{BAMPS,Uphoff:2014cba} and study the impact of the jet-medium coupling $\kappa=\kappa(E^2,T)$ 
derived by CUJET3.0 \cite{Xu:2014tda}.

We show that both medium backgrounds lead to surprisingly similar results. However, 
the background flow fields need to be included in both scenarios to enhance the 
high-$p_T$ elliptic flow which is otherwise too small. Applying the jet-medium coupling 
$\kappa=\kappa(E^2,T)$ finally leads to a reproduction of the high-$p_T$ $v_2$-data 
within measured error bars. 

\begin{figure*}[t]
\begin{center}
\includegraphics*[width=14.5cm]{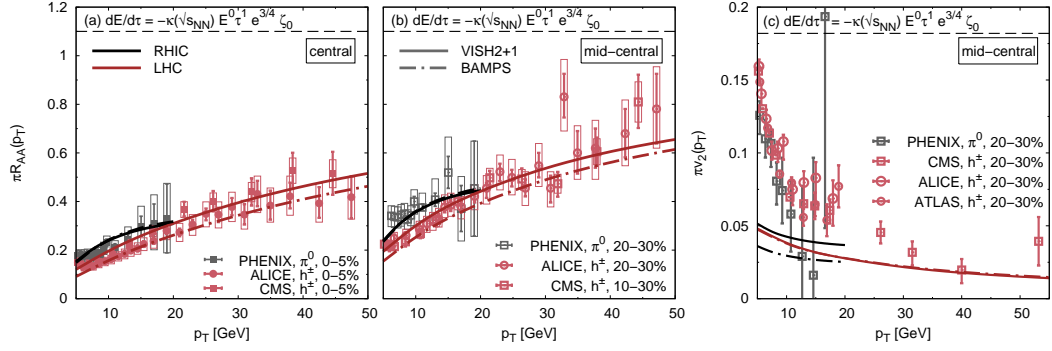}
\caption{The pion nuclear modification factor for central (left panel) and mid-central 
(middle panel) collisions at RHIC (black) and LHC (red) as well as the high-$p_T$ pion
elliptic flow for mid-central events (right panel). The measured data 
\cite{data1,data2,data3,data4} are compared to a pQCD-based energy loss 
$dE/d\tau=\kappa(\sqrt{s_{NN}}) E^0 \tau^1 e^{3/4} \zeta_{0}$ that includes 
jet-energy loss fluctuations ($\zeta_0$). The jet-medium coupling depends 
on the collision energy [$\kappa=\kappa(\sqrt{s_{NN}})$] and the medium
background is either described applying the hydrodynamic code VISH2+1 \cite{VISH2+1} 
(solid lines) or the parton cascade BAMPS \cite{BAMPS} (dashed-dotted lines).
The background flow $\Gamma_f$ is not included here.
\vspace*{-0.4cm}}
\label{Fig01}
\end{center}
\end{figure*}

With this, we contrast two completely different background models and demonstrate 
the importance of the background flow fields and the jet-medium coupling 
for the correct description of the measured jet observables.

\section{The Jet-Energy Loss Model BBMG}

The jet-energy loss model used in this letter (for convenience referred to as BBMG model) 
is based on the generic ansatz \cite{Betz:2014cza,Betz:2012qq}
\begin{eqnarray}
\frac{dE}{d\tau}= 
-\kappa\,  E^a(\tau) \, \tau^{z} \, e^{c=(2+z-a)/4} \, \zeta_q \, \Gamma_f ,
\label{Eq1}
\end{eqnarray}
with the jet-energy $E$, the path-length $\tau$, and the energy density of 
the background medium $e$. 

In case of a radiative perturbative QCD (pQCD) energy-loss description used here, the explicit form of 
Eq.\ (\ref{Eq1}) is \cite{Betz:2014cza,Betz:2012qq}
\begin{eqnarray}
\frac{dE}{d\tau}= 
-\kappa\,  E^0(\tau) \, \tau^{1} \, e^{3/4} \, \zeta_q \, \Gamma_f .
\label{Eq2}
\end{eqnarray}
Jet-energy loss fluctuations are included via the distribution
$f_q(\zeta_q)= \frac{(1 + q)}{(q+2)^{1+q}} (q + 2- \zeta_q)^q $ which allows
for an easy interpolation between non-fluctuating ($\zeta_{q=-1}=1$) distributions 
and those ones increasingly skewed towards small $\zeta_{q>-1} < 1$. Unless
mentioned otherwise, the jet-energy loss fluctions are included with $q=0$ 
\cite{Betz:2014cza}.

The background flow fields are incorporated via the flow factor 
$\Gamma_f=\gamma_f [1 - v_f \cos(\phi_{\rm jet} - \phi_{\rm flow})]$ with 
the background flow velocities $v_f$ given by VISH2+1 \cite{VISH2+1} or
BAMPS \cite{BAMPS} and the $\gamma$-factor $\gamma_f = 1/\sqrt{1-v_f^2}$ 
\cite{Liu:2006he,Baier:2006pt,Renk:2005ta,Armesto:2004vz}.
$\phi_{\rm jet}$ is the jet angle w.r.t.\ the reaction plane and 
$\phi_{\rm flow}=\phi_{\rm flow}(\vec{x},t)$ is the corresponding local azimuthal
angle of the background flow fields.

Initially, the jets are distributed according to a transverse initial profile 
given by the bulk flow fields of VISH2+1 and BAMPS \cite{VISH2+1,BAMPS}.

Besides the two background media, we also contrast a jet-medium coupling 
$\kappa$ that depends only on the collision energy $\kappa=\kappa(\sqrt{s_{NN}})$ 
with the CUJET3.0 jet-medium coupling $\kappa=\kappa(E^2,T)$ which depends on 
the jet energy, the local temperature, and includes possible non-perturbative effects 
around the phase transition as in Ref.\ \cite{Xu:2014tda}. 
The DGLV \cite{DGLV} jet-medium coupling was generalized in Ref.\ \cite{Xu:2014tda}
to be of the form
\begin{eqnarray}
\kappa(E^2,T) &=& \alpha_S^2(E^2)\chi_T\left(f_E^2+f_E^2 f_M^2 \mu^2/E^2\right)\nonumber\\
&& - (1-\chi_T)(f_M^2 + f_E^2 f_M^2\mu^2/E^2)
\,.
\label{Eq3}
\end{eqnarray}
Please note that Eq.\ (\ref{Eq3}) is qualitatively similar to CUJET3.0 as the running 
coupling constant there a function of momentum transfer $\alpha(Q^2)$ while we  
assume that $Q^2=E^2$. The above expression includes 
\begin{itemize}
\item[(1)] a running coupling effect via $\alpha_S(E^2)=1/[c+9/4\pi\,\log(E^2/T_c^2)]$ with
$c=1.05$, 
\item[(2)] the Polyakov-loop suppression of the color-electric scattering \cite{Hidaka:2008dr}
via $\chi_T = c_q L + c_g L^2$ with the pre-factors $c_q = (10.5 N_f)/(10.5 N_f + 16)$ 
for quarks and $c_g = 16/(10.5 N_f + 16)$ for gluons. Here, we consider $N_f = 3$. 
$L(T) = [\frac{1}{2} + \frac{1}{2} \rm{Tanh}[7.69(T - 0.0726)]]^{10}$ is a fit to 
lattice QCD \cite{Bazavov:2009zn,Borsanyi:2010bp}, as in Ref.\ \cite{Xu:2014tda}.
\item[(3)] and a model of near-critical $T_c$ enhancement of scattering due to 
emergent magnetic monopoles. The functions $f_E(T)$ and $f_M(T)$ are also fits 
to lattice QCD \cite{Nakamura:2003pu}. The electric and magnetic screening masses 
are given by $\mu_{E,M}(T) = f_{E,M}(T)\mu(T)$ with 
the Debye screening mass $\mu^2(T) = \sqrt{4\pi\alpha_s(\mu^2)}T\sqrt{1+N_f/6}$.
The functions $f_{E,M}(T)$ can be re-written to $f_E(T) = \sqrt{\chi_T}$ and
$f_M(T)=0.3 \mu(T)/(T\sqrt{1+N_f/6})$ \cite{Xu:2014tda}.
\end{itemize}
This jet-medium coupling decreases with the temperature of the background medium and thus
shows an effective running with collision energy. 

\begin{figure*}[t]
\begin{center}
\includegraphics*[width=14.5cm]{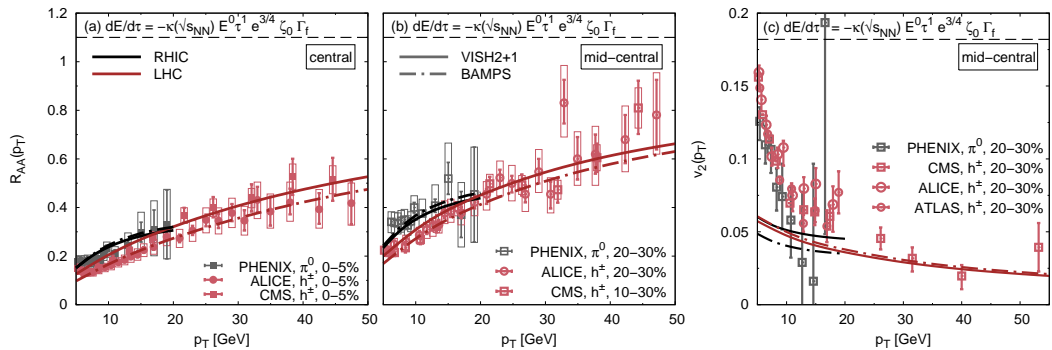}
\caption{The pion nuclear modification factor for central (left panel) and mid-central 
(middle panel) collisions at RHIC (black) and LHC (red) as well as the high-$p_T$ pion
elliptic flow for mid-central events (right panel). The measured data 
\cite{data1,data2,data3,data4} are compared to a pQCD-based energy loss 
$dE/d\tau=\kappa(\sqrt{s_{NN}}) E^0 \tau^1 e^{3/4} \zeta_{0} \Gamma_f$, including jet-energy loss fluctuations 
($\zeta_0$) and transverse flow fields ($\Gamma_f$), for a jet-medium coupling that depends 
on the collision energy [$\kappa=\kappa(\sqrt{s_{NN}})$] applying the hydrodynamic backgrounds of VISH2+1 \cite{VISH2+1} 
(solid lines) and the parton cascade BAMPS \cite{BAMPS} (dashed-dotted lines).
\vspace*{-0.4cm}}
\label{Fig02}
\end{center}
\end{figure*}

\section{Results and Discussion}

Fig.\ \ref{Fig01} shows the pion nuclear modification factor ($R_{AA}$) 
for central (left panel) and mid-central (middle panel) collisions at RHIC 
(black) and LHC (red) as well as the high-$p_T$ pion elliptic flow ($v_2$) for 
mid-central events (right panel). The measured data \cite{data1,data2,data3,data4} 
are compared to the pQCD-based energy loss of Eq.\ (\ref{Eq2}) excluding
the flow fields $\Gamma_f$, $dE/d\tau=\kappa(\sqrt{s_{NN}}) E^0 \tau^1 e^{3/4} \zeta_{0}$,
for the hydrodynamic backgrounds of VISH2+1 \cite{VISH2+1} 
(solid lines) and the parton cascade BAMPS \cite{BAMPS} (dashed-dotted lines).
Jet-energy loss fluctuations are considered via $\zeta_0$. The jet-medium 
coupling in Fig.\ \ref{Fig01} depends on the collision energy 
[$\kappa=\kappa(\sqrt{s_{NN}})$]. 

Within the present error bars, both the central and the mid-central pion nuclear
modification factor can be described using this pQCD ansatz without the 
background flow fields. However, the  high-$p_T$ $v_2$-problem \cite{Betz:2014cza,Betz:2012qq}
becomes obvious. The right panel of Fig.\ \ref{Fig01} shows that the 
high-$p_T$ elliptic flow is below the measured data.

Including the background flow fields via $\Gamma_f$ in
Fig.\ \ref{Fig02} leads to a significant increase of the high-$p_T$ elliptic 
flow while the pion nuclear modification factor is only affected marginally. 

Fig.\ \ref{Fig02} reveals the strong influence of the background flow fields 
on the high-$p_T$ elliptic flow.

Besides this, Figs.\ \ref{Fig01} and \ref{Fig02} demonstrate a surprising 
similarity between the results based on a medium described by viscous hydrodynamics
\cite{VISH2+1} and the parton cascade BAMPS \cite{BAMPS}. This similarity 
cannot be expected a priori as the two background media are quite
different: While the hydrodynamic description of VISH2+1 assumes an 
equilibrated system, the parton cascade BAMPS also includes non-equilibrium effects
in the bulk medium evolution. However, since those effects are small, a
temperature can be defined after a very short initial time $t_0$. In this work, 
we use $t_0=0.3$~fm at RHIC and $t_0=0.2$~fm at LHC energies.

\begin{figure*}[t]
\begin{center}
\includegraphics*[width=14.5cm]{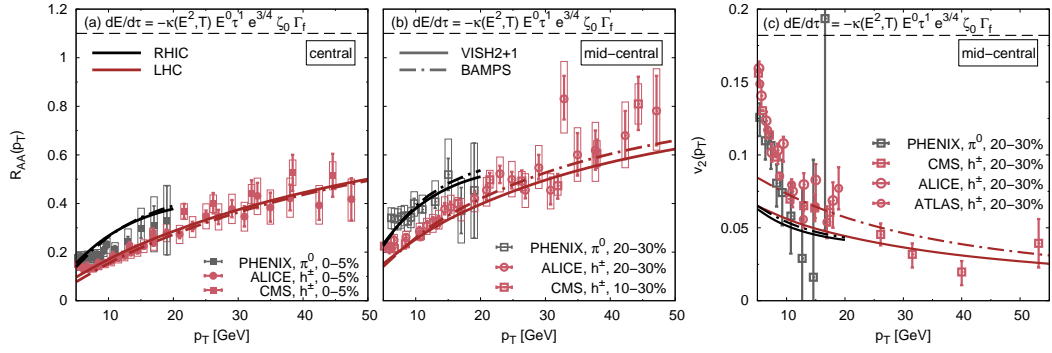}
\caption{The pion nuclear modification factor for central (left panel) and mid-central 
(middle panel) collisions at RHIC (black) and LHC (red) as well as the high-$p_T$ pion
elliptic flow for mid-central events (right panel). The measured data 
\cite{data1,data2,data3,data4} are compared to a pQCD-based energy loss 
$dE/d\tau=\kappa(E^2,T) E^0 \tau^1 e^{3/4} \zeta_{0} \Gamma_f$, including jet-energy loss fluctuations 
($\zeta_0$) and transverse flow fields ($\Gamma_f$), for a jet-medium coupling that depends 
on the energy of the jet, the temperature of the medium, and non-equilibrium
effects around the phase transition [$\kappa=\kappa(E^2,T)$]. The medium background
is either described applying the hydrodynamic backgrounds of VISH2+1 \cite{VISH2+1} 
(solid lines) and the parton cascade BAMPS \cite{BAMPS} (dashed-dotted lines). 
\vspace*{-0.4cm}}
\label{Fig03}
\end{center}
\end{figure*}

In a third step, we include the jet-medium coupling $\kappa=\kappa(E^2,T)$
given by Eq.(\ref{Eq3}) \cite{Xu:2014tda} in our jet-energy loss approach.
The result is shown in Fig.\ \ref{Fig03}, again for the hydrodynamic background 
VISH2+1 (solid lines) and a medium determined via the parton cascade BAMPS 
(dashed-dotted lines). 
As in Figs.\ \ref{Fig01} and \ref{Fig02}, the pion nuclear modification factor is well 
described both at RHIC and LHC. However, the high-$p_T$ elliptic flow increases significantly
below $p_T \sim 20$~GeV, especially for the BAMPS background which already includes 
non-equilibrium effects through microscopic, non-equilibrium transport 
calculations \cite{BAMPS,Uphoff:2014cba}. 

Fig.\ \ref{Fig03} demonstrates that the jet-medium coupling $\kappa=\kappa(E^2,T)$
suggested by CUJET3.0 \cite{Xu:2014tda} can solve
the high-$p_T$ $v_2$-problem. However, the background medium considered does
play an important role for the description of the high-$p_T$ elliptic flow. 
Please note that the initial conditions studied here are non-fluctuating, 
i.e.\ neglect event-by-event short-scale inhomogeneities. The effect of 
event-by-event fluctuations will be studied elsewhere \cite{Betz:2016}.

\begin{figure*}[t]
\begin{center}
\includegraphics*[width=14.5cm]{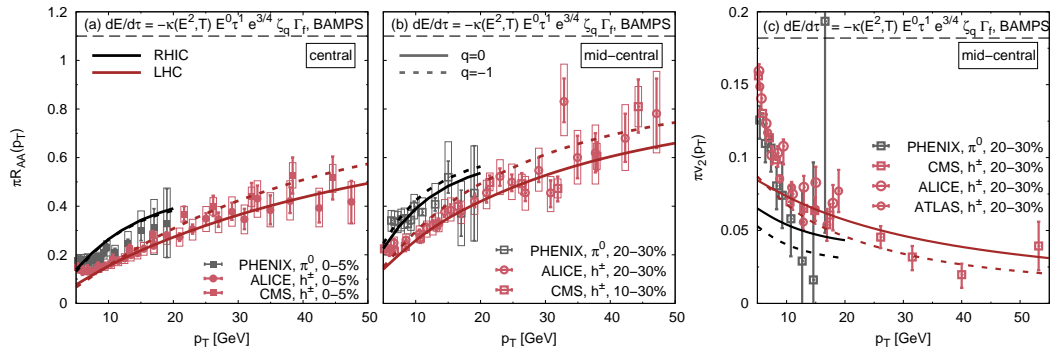}
\caption{The pion nuclear modification factor for central (left panel) and mid-central 
(middle panel) collisions at RHIC (black) and LHC (red) as well as the high-$p_T$ pion
elliptic flow for mid-central events (right panel). The measured data 
\cite{data1,data2,data3,data4} are compared to a pQCD-based energy loss with transverse
medium flow, $dE/d\tau=\kappa(E^2,T) E^0 \tau^1 e^{3/4} \zeta_{q} \Gamma_f$. Jet-energy loss fluctuations 
are included ($q=0$) or excluded ($q=-1$). The jet-medium coupling depends 
on the energy of the jet, the temperature of the medium, and non-equilibrium
effects around the phase transition [$\kappa=\kappa(E^2,T)$]. The background medium
is described by the parton cascade BAMPS \cite{BAMPS}. 
\vspace*{-0.4cm}}
\label{Fig04}
\end{center}
\end{figure*}

To further investigate the influence of the parton cascade medium, we varied 
the jet-energy loss fluctuations. The results are shown in Fig.\ \ref{Fig04}. 
As one can see, the slope of both the $R_{AA}$ and the high-$p_T$ $v_2$ at LHC 
energies changes when considering non-fluctuating ($\zeta_{q=-1}=1$) jet-energy 
loss distributions. To be more precise, the results get closer to the 
slope of the measured data. This is in contrast to results previously obtained 
with the pQCD-based jet-energy loss ansatz \cite{Betz:2014cza} 
for the VISH2+1 background \cite{VISH2+1}. In Ref.\ \cite{Betz:2014cza} 
the slope of both the $R_{AA}$ and the high-$p_T$ $v_2$ did not change significantly
when changing the jet-energy loss fluctuations. In particular, the results of 
Ref.\ \cite{Betz:2014cza} for the elliptic flow at $p_T>20$~GeV coincided 
for various fluctuation distributions.

This result strengthens the observation that the background medium considered
plays an important role for the correct description of the jet observables. 
However, while the nuclear modification factor is less influenced by the 
background medium, the impact of the background medium and background flow 
on the high-$p_T$ elliptic flow is quite significant.

\section{Conclusions}

We compared the measured data on the nuclear modification factor and the 
high-$p_T$ elliptic flow at RHIC and LHC energies to results 
obtained by the pQCD-based jet-energy loss model BBMG. We contrasted results
obtained via a hydrodynamic background (VISH2+1) \cite{VISH2+1} with results
based on the parton cascade BAMPS \cite{BAMPS,Uphoff:2014cba}. We showed that the results for 
both medium backgrounds exhibit a remarkable similarity, especially for the
pion nuclear modification factor. We demonstrated that 
the background medium and background flow strongly influence the high-$p_T$ $v_2$.
We found that for event-averaged or non-fluctuating initial
conditions, studied here, the simultaneous description of the pion nuclear 
modification factor and high-$p_T$ elliptic flow requires to consider both 
the background flow fields and a jet-medium coupling that depends on the energy of the
jet, the temperature of the medium, and non-equilibrium effects
around the phase transition \cite{Xu:2014tda}. 

\vspace*{1cm}\section{Acknowledgement}
We thank J.\ Noronha, J.\ Noronha-Hostler, and J.\ Xu for helpful discussions 
as well as U.\ Heinz and C.\ Shen for making their hydrodynamic field grids 
available. This work was supported 
through the Bundesministerium f\"ur Bildung und Forschung, the Helmholtz 
International Centre for FAIR within the framework 
of the LOEWE program (Landesoffensive zur Entwicklung 
Wissenschaftlich-\"Okonomischer Exzellenz) launched by the State of Hesse, 
the US-DOE Nuclear Science Grant No.\ DE-AC02-05CH11231 within the framework 
of the JET Topical Collaboration, the US-DOE Nuclear Science Grant No.\
DE-FG02-93ER40764, and IPP/CCNU, Wuhan. Numerical computations have been performed 
at the Center for Scientific Computing (CSC).

\end{document}